\documentclass[aps,prd,twocolumn, showpacs,nofootinbib]{revtex4}
\usepackage{amsmath}
\usepackage{graphicx}
\usepackage{dcolumn}
\usepackage{bm}
\usepackage{amssymb}
\usepackage{latexsym}

\newcommand{\no}{\nonumber}

\newcommand{\pa}{\partial}
\newcommand{\del}{\delta}
\newcommand{\eps}{\epsilon}

{\bf }

\newcommand{\half}{\frac{1}{2}}

\newcommand{\sig}{\sigma}
\newcommand{\Del}{\Delta}

\begin{document}

\title{Quantum speed limit for a relativistic  electron in the noncommutative phase space}

\author{Kang Wang ${}^a$}

\author{Yu-Fei Zhang ${}^a$}

\author{Qing Wang ${}^b$}

\author{Zheng-Wen Long ${}^c$}





\author{Jian Jing ${}^{a}$}
\email{jingjian@mail.buct.edu.cn}

\affiliation{${}^a$ Department of Physics and Electronic, School of
Science, Beijing University of Chemical Technology, Beijing 100029,
P.R. China,}

\affiliation{${}^{b}$ College  of Physics and Technology, XinJiang University, Urumqi 830046, P.R.China}

\affiliation{$^{c}$ Department of Physics, GuiZhou University, GuiYang,
550025}

\begin{abstract}
The influence of the noncommutativity on the average speed of a relativistic electron  interacting with a uniform magnetic field within the minimum evolution time is investigated. We find that it is possible for the wave packet of  the electron to travel faster than the speed of light in  vacuum because of the noncommutativity. 

\end{abstract}

{\pacs {11.10.Ef, 03.65.Pm, 03.65.Ge}}

\maketitle


\section{Introduction}

The time of a quantum system evolving from an initial state to the orthogonal final state is of great importance in the field of quantum computation, quantum control and quantum metrology. In fact, it has attracted  attention  for a long time \cite{mt}. It is proven in Refs. \cite{fvva} that the minimum  time $T_{\rm min}$ for a quantum system starting from an initial state
to an orthogonal final state is given by $T_{\rm min} = \frac{\pi \hbar}{2 \Del E}$, in which $\Del E$ is the energy variance, defined by  $\Del E = \sqrt {\langle \psi |H ^2| \psi \rangle -\langle \psi |H |\psi \rangle ^2}$, with $H$ being the Hamiltonian of the system. However, a somewhat different result is presented in Ref. \cite{ml}. The authors of this paper show  that the minimum time is determined by the average energy $\bar E= \langle \psi|H |\psi \rangle$. They show that the minimum time for the quantum system evolving from an initial state to the orthogonal one is $T_{\rm min} = \frac{\pi \hbar}{2(\bar E -E_0)}$, where $E_0$ is the lowest energy of the state which participates in the superposition. Evidently, the results of Ref. \cite{fvva} and \cite{ml} will be equivalent if the condition $\Del E = \bar E -E_0$ is satisfied.  This condition 
can be satisfied  by superposing two steady states homogeneously.
According to the results in Refs. \cite{fvva, ml}, one naturally assumes that the minimum time should be
given by $T_{\rm min} = {\rm Max} \{\frac{\pi \hbar}{2 \Del E}, \ \frac{\pi \hbar}{2(\bar E -E_0)}\}$ \cite{glm}.
A unified bound which contains both $\Delta E$ and $\bar E$ is considered \cite{lt}.

An interesting connection between the  speed of the quantum state evolving in Hilbert space and the speed of the electron wave packet traveling  in spatial  space is constructed in a recent paper \cite{vd}. In this paper, the authors choose a relativistic quantum mechanical model, namely, a relativistic electron interacting with a uniform magnetic field. The authors find that the average speed of the electron wave packet moving in the radial direction during the interval in which the quantum state evolving from an initial state to the orthogonal final state in Hilbert space is less than the speed of light in vacuum, regardless of the intensity of the magnetic field one applies. It seems that  Lorentz invariance is not violated  in this relativistic quantum mechanical model. Lorentz invariance would be violated in the non-relativistic limit of this model since the average speed of the electron wave packet in the radial direction exceeded the speed of light in the vacuum provided the intensity of the magnetic field is  strong enough.

On the other hand,  the noncommutativity has attracted much attention due to superstring theories in the past years \cite{string1, string2, string3, string4}. There are tremendous papers studying quantum field theories in noncommutative space \cite{nqf1,nqf2,nqf3,nqf4,nqf5,nqf6}.
Non-relativistic noncommutative quantum mechanics, such as noncommutative harmonic oscillator, noncommutative Landau problem have also been studied extensively \cite{ncqm1,ncqm2,ncqm3,ncqm4,ncqm5,ncqm6,ncqm7}.
The relativistic quantum mechanical models on noncommutative space are also investigated since the work of \cite{ndha}. Interestingly, it is found that the noncommutativity even has a profound relationship with JC model in quantum optics context \cite{optics}. Some geometrical phases in noncommutative relativistic quantum theory are studied  Refs. \cite{ma1, ma2, ma3} recently.
In Ref. \cite{jackiw}, the authors show that due to the noncommutativity, Lorentz invariance will be violated in the noncommutative quantum electro-dynamics (QED) since the electromagnetic wave travels in  different speed along different directions at the presence of a background magnetic field. A similar result is also obtained in Ref. \cite{cai}.  The work of \cite{jackiw, cai} may afford a possible method to detect the spatial noncommutativity via the Michelson-Morley-type experiment. From the quantum point of view electromagnetic field is photons, which is massless.  An interesting question is: does a massive particle can travel faster than light in vacuum due to the noncommutativity? In Refs. \cite{Deriglazov1, Deriglazov2, Deriglazov3}, the authors consider this question semi-classically from both noncommutative and gravitational  points of view.  In this paper, we will consider this question quantum mechanically. 

It is well-known that Dirac equation is one of the successful relativistic quantum mechanics theories which describes spin-$\half$ particles. It accommodates the quantum mechanics and the special relativity perfectly. The achievements of Dirac equation are various, including the fine structure of the Hydrogen atoms, the prediction of anti-particles. It also affords a natural description of electron spin. Because of these, the authors of Ref. \cite{vd} connect the evolution of the internal states which are determined by Dirac equation and the spatial motion of the electron wave packet to test whether Lorentz invariance is violated.
In this paper, we shall investigate the noncommutative corrections of Ref. \cite{vd},  
i.e., we shall investigate whether the speed of  electrons  can exceed the speed of light in vacuum because of the noncommutativity.  

The organization of the present paper is as follows: in the next section, we shall start our studies from analysing the two-dimensional case. We shall show that 
the speed of the wave packet of an electron can exceed the speed of light in vacuum provided the intensity of the magnetic field is strong enough. Then, in the section III, we generalize our studies to the three-dimension case. We find that for a specific noncommutative configuration, the similar conclusions can also be obtained in three-dimensional case. Some conclusions and remarks will be given in the last section.

\section{Two-dimensional case}

It is well known that Dirac equation describes  relativistic spin-$\half$ particles. It can be written in the Schrodinger equation form
\begin{equation}
i \hbar \frac{\pa \psi}{\pa t} =H \psi, \label{ste}
\end{equation}
in which $\psi$ is a four-component wave function $\psi = (\psi_1, \ \psi_2, \ \psi_3, \psi_4)^T$, $H$ is the Hamiltonian. It is given by \footnote{The summation convention is applied in this paper and the capital Latin Letters $I,\ J, \ K$ run from $1$ to $3$ while the lowercase $i, \ j$ run from $1$ to $2$. }
\begin{equation}
H = c \alpha_I p_I  + m c^2 \beta \label{3dh}
\end{equation}
where $m$ is the rest mass, $\alpha_I, \ \beta$ are Dirac matrices. They satisfy the following anti-commutation relations
\begin{equation}
\{\alpha_I, \ \alpha_J \} = 2 \del_{IJ}, \quad \{\alpha,\ \beta \}=0.
\end{equation}
The Dirac matrices $\alpha_I, \ \beta$ can be realized by the Pauli matrices and the $2 \times 2$ identity matrix $I$, i.e., $$\alpha_I = \left (\begin{array}{cc}0& \sig_I \\ \sig_I & 0 \end{array} \right), \ \beta= \left( \begin{array}{cc} I &0 \\ 0&-I \end{array} \right ).$$

In two dimensional space, the Dirac matrices $\alpha_I$ and $\beta$ are simplified to the Pauli matrices, i.e., $\alpha_1 = \sig_1, \ \alpha_2 = \sig_2$ and $\beta = \sig_3$. The wave function reduces to a two-component one $\psi= (\phi, \ \chi)^T$. Thus, the Hamiltonian of Dirac equation in two-dimensional space is
\begin{equation}
H= c \sig_i p_i + mc^2 \sig_3. \label{tdh}
\end{equation}
In the presence of a magnetic field, the coupling between the electric charge and the magnetic field should be introduced by the minimal substitution $$p_i \to p_i + e A_i$$ in the Hamiltonian (\ref{tdh}) with $e$ being the absolute value of electron charge. Thus, the Dirac equation reduces to the form
\begin{equation}
i \hbar \frac{\pa }{\pa t} \left( \begin{array}{c} \phi \\ \chi \end{array} \right) =H  \left( \begin{array}{c} \phi \\ \chi \end{array} \right), \label{2dde}
\end{equation}
where the Hamiltonian (\ref{tdh}) becomes
\begin{equation}
H = c\sig_i (p_i + e A_i) + m c^2 \sig_3. \label{h10}
\end{equation}
The explicit expression of Hamiltonian (\ref{tdh}) is
\begin{equation}
H=\left ( \begin{array} {cc} H_{11}& H_{12} \\ H_{21}& H_{22} \end{array} \right)
\end{equation}
where
\begin{eqnarray}
H_{11}&=& - H_{22}=  mc^2, \no \\
H_{12}&=& H_{21} ^* = c(p_1 - i p_2) + e c (A_1 - i A_2).
\end{eqnarray}
Thus, the eigenvalue equation $H \psi = E \psi$ turns out to be
\begin{eqnarray}
(E- m c^2) \phi &=& c[ p_1 - i p_2 + e  (A_1 - i A_2)] \chi, \no \\
(E+ m c^2) \chi &=& c[ p_1 + i p_2 + e  (A_1 + i A_2)] \phi.  \label{cnee}
\end{eqnarray}
Obviously,  $\phi$ is the larger  component while $\chi$ is the smaller one since it will tend to zero if the non-relativistic limit
$
E=mc^2 + \varepsilon \ (\eps \ll mc^2)
$ is taken.

In the two-dimensional case, the noncommutative phase space is described by the algebraic relations \footnote{We only consider the noncommutativities among coordinates and momenta so as to guaranteeing the unitarity.} \cite{bertolami}
\begin{equation}
[\hat x_i, \ \hat x_j ] = i \theta_{ij}, \quad [\hat p_i, \ \hat p_j ] = i \eta_{ij}, \quad [\hat x_i, \ \hat p_j] = i \hbar_{\rm eff} \del_{ij},
\label{ncps}
\end{equation}
where $\theta_{ij} = \theta \eps_{ij}, \ \eta_{ij} = \eta \eps_{ij}$, $\hbar_{\rm eff} =(1+ \frac{\theta \eta}{4 \hbar^2}) \hbar$ with $\eps_{ij}$ being the 2-dimensional anti-symmetric tensor,  $\theta$ and $ \eta$ being two   real parameters. It is assumed that both of these two parameters are small.
The upper bound of these two parameters is estimated in some literatures \cite{bertolami}.

In the general description of noncommutative quantum mechanics, it is assumed that the dynamical equations take the same form as their commutative counterparts, however, variables in Hamiltonian are replaced by the corresponding noncommutative ones. Therefore, Dirac equation in noncommutative phase space  (\ref{ncps}) takes the same form as in (\ref{2dde}) except for the variables in Hamiltonian (\ref{h10}) are replaced by the noncommutative variables which satisfied  (\ref{ncps}).

We choose the symmetric gauge  \cite{zhangjz}
$$\hat A_i = - \frac{B}{2} \eps_{ij} \hat x_j$$
and map the noncommutative variables onto the commutative ones which satisfy the standard Heisenberg algebra
\begin{equation}
[x_i, \ x_j]=[p_i, \ p_j]=0, \quad [x_i, \ p_j] = i \hbar \del_{ij}.
\end{equation}
It is straightforward to check that the map from the noncommutative variables onto the commutative ones can be realized by
\begin{equation}
\hat x_i = x_i - \frac{\theta}{2 \hbar} \eps_{ij} p_j, \quad \hat p_i = p_i + \frac{\eta}{2 \hbar} \eps_{ij} x_j. \label{cvs}
\end{equation}

In terms of commutative variables $(x_i, \ p_i)$, we rewrite the equation (\ref{cnee}) as
\begin{eqnarray}
(E- m c^2) \phi &=& c[ (1- \frac{e \theta B}{4 \hbar}) (p_1 -i p_2) \no \\ &&+ i (\frac{ \eta}{2 \hbar} - \frac{eB}{2}) (x_1 - i x_2)] \chi \label{eeee1} \\
(E+ m c^2) \chi &=& c[ (1- \frac{e \theta B}{4 \hbar}) (p_1 + i p_2) \no \\ &&- i (\frac{ \eta}{2 \hbar} - \frac{eB}{2}) (x_1 + i x_2)] \phi. \label{eeee}
\end{eqnarray}
Multiplying $(E+ m c^2)$ on both sides of equation (\ref{eeee1}) from left, we get
\begin{eqnarray}
(E^2 - m^2 c^4) \phi &=& c^2 \bigg[ (1- \frac{e \theta B}{4 \hbar})^2  p_i^2 + (\frac{\eta}{2 \hbar} - \frac{eB}{2})^2 x_i ^2 \no \\ && - 2 (\frac{\eta}{2 \hbar} - \frac{e B}{2})(1- \frac{e\theta B}{4 \hbar})(L_z + \hbar)  \bigg] \phi. \label{cde}
\end{eqnarray}
Up to a constant, the bracket on the right-hand side of the above equation is proportional to the Hamiltonian of a two-dimensional harmonic oscillator with a rotational term. 
The effective mass and the frequency  are
\begin{eqnarray}
M &=& \frac{m}{ (1 - \frac{e \theta B}{4 \hbar})^2}, \label{mass} \\
\Omega &=& \frac{1}{m} \big |(1- \frac{e \theta B}{4 \hbar}) (\frac{e B}{2}- \frac{ \eta}{2 \hbar}) \big |.  \label{frequency}
\end{eqnarray}

The eigenvalues and eigenfunctions of the  equation (\ref{cde}) can be solved directly. They are \footnote{We only consider the positive energy sector in this paper.}\cite{yoshioka}
\begin{eqnarray}
E_{n,m_l}= \pm \sqrt{m^2 c^4 + 2 mc^2 \hbar \Omega (n+m_l+2)}
\end{eqnarray}
and
\begin{eqnarray} \label{esnrl}
\phi_{n,m_l} (r, \varphi) &=& \frac{{(-1)}^{\frac{n- |m_l|}{2}} (\frac{n- |m_l|}{2})!}{\sqrt {\pi (\frac{n+|m_l|}{2})! (\frac{n- |m_l|}{2})!} } \label{efs} \\
&& \times \alpha (\alpha r) ^ {|m_l|} L_{(\frac{n-|m_l|}{2})} ^{|m_l|} (\alpha^2 r^2) e ^{- \half \alpha ^2 r^2} e ^{i m_l \varphi}. \no
\end{eqnarray}
In which $n$, $m_l$ take values $n=0,1,2,\cdots$, $m_l =- n, -n+2, \cdots, n-2, n$, $ L_{(\frac{n-|m_l|}{2})} ^{|m_l|} $ is the generalized Laguerre polynomials, $\alpha$ is a dimensionless parameter. It is defined by $\alpha = \sqrt {\frac{M\Omega}{\hbar}}.$
The smaller component of the eigenfunction  is determined by (\ref{eeee}) once the larger component is obtained.

We shall prepare a specific superposition state for further discussions.
In order to make a comparison with its commutative counterpart \cite{vd}, we shall superpose two steady states. We choose the larger components of these two steady states as
$\phi_{0,0} \ {\rm and} \  \phi_{2,0}$
respectively.
Then, after a direct calculation, we find the corresponding two smaller components. They  are
\begin{eqnarray}
\chi_{0,0} &=& \frac{2 i c \hbar \alpha (1- \frac{e \theta B}{4 \hbar}) \phi_{1,1}}{ E_{0,0} + m c^2},  \no \\
\chi_{2,0} &=& \frac{ 2 \sqrt 2 i c\hbar \alpha( 1- \frac{e \theta B}{4 \hbar}) \phi_{3,1}}{E_{2,0} + m c^2}.
\end{eqnarray}
Thus, two steady states which we shall superpose are
\begin{equation}
\Phi_{0,0} (r, \varphi,t)= N_{0,0} \left ( \begin{array}{c} \phi_{0,0} \\ \frac{2 i c \hbar \alpha (1- \frac{e \theta B}{4\hbar}) \phi_{1,1}}{E_{0,0} + mc^2} \end{array} \right ) e^{- i \frac{E_{0,0}}{\hbar} t} \label{2ef00}
\end{equation}
and
\begin{equation}
\Phi_{2,0} (r, \varphi,t)= N_{2,0} \left (\begin{array}{c} \phi_{2,0} \\ \frac{2 \sqrt 2 i c \hbar \alpha (1- \frac{e \theta B}{4 \hbar}) \phi_{3,1} }{E_{2,0} + mc^2} \end{array} \right)e^{- i \frac{E_{2,0}}{\hbar} t} ,\label{2ef20}
\end{equation}
with $N_{0,0}$ and $N_{2,0}$ being two normalization constants
\begin{eqnarray}
N_{0,0} =\frac{E_{0,0} + m c^2}{\sqrt{(E_{0,0} + m c^2)^2 + 4 \hbar^2 c^2 \alpha^2 (1- \frac{e \theta B}{4 \hbar})^2}} , \no \\
N_{2,0} =\frac{E_{2,0} + m c^2}{\sqrt{(E_{2,0} + m c^2)^2 + 8 \hbar^2 c^2 \alpha^2 (1- \frac{e \theta B}{4 \hbar})^2}}. \label{nconst}
\end{eqnarray}

For the sake of  avoiding the ambiguity of whether the minimum time is determined by the energy variance $\Del E$ or the average energy $\bar E$,  we superpose two eigenfunctions (\ref{2ef00}, \ref{2ef20}) homogeneously. Thus, the superposition state we prepared is
\begin{equation}
\Psi(r,\varphi,t) = \frac{1}{\sqrt 2} \big ( \Phi_{0,0}(r, \varphi,t) + \Phi_{2,0}(r, \varphi,t) \big). \label{state}
\end{equation}

We consider the extreme case $B \to \infty$ \cite{vd}.
It means that the magnetic field is strong enough so one can neglect the rest energy $m c^2$.   In this case, the eigenvalues $E_{0,0}$ and $E_{2,0}$ as well as normalization constant (\ref{nconst}) reduce to
\begin{equation}
E_{0,0} = 2 \hbar c \alpha \big| 1- \frac{e \theta B}{4 \hbar} \big|, \ E_{2,0} = 2 \sqrt 2 \hbar c \alpha \big |1- \frac{e \theta B}{4 \hbar} \big | \label{evs}
\end{equation}
and
\begin{equation}
N_{0,0}= N_{2,0} = \frac{1}{\sqrt 2}.
\end{equation}

Now, we are ready to calculate the minimum time for the state (\ref{state}) evolving from the initial state $\Psi(r, \varphi, 0)$ to the  orthogonal final one $\Psi(r, \varphi, T_{min})$. Substituting the eigenvalues (\ref{evs}) into $T_{\rm min} = \frac{\pi \hbar}{2(\bar E -E_0)}$, we get the minimum time. It is
\begin{equation}
T_{\rm min} = \frac{\pi \hbar}{\bar E - E_{0,0}} = \frac{\pi}{2 (\sqrt{ 2}-1) c \alpha \big |1- \frac{e \theta B}{4 \hbar} \big| }.  \label{mt}
\end{equation}

The displacement of the wave packet moving in the radial direction during this period of time can be obtained by substituting the wave functions (\ref{state}) into the expression
\begin{eqnarray}
\big| \Del r \big| &=& \bigg |\langle \Psi (T_{min}) |  r|  \Psi(T_{min})\rangle - \langle \Psi(0) |r| \Psi(0) \rangle \bigg | \no \\
&=&\bigg |\langle \Psi_{0,0}|r | \Psi_{2,0} \rangle [\cos  (\Omega^\prime T_{\rm min})-1] \bigg |,
\end{eqnarray}
where $\Omega ^\prime = \frac{E_{2,0} -E_{0,0}}{\hbar}= 2(\sqrt 2 -1)c \alpha \big |1- \frac{e\theta B}{4 \hbar} \big |$.
After some calculations, we arrive at
\begin{eqnarray}
\Del r= \frac{\sqrt \pi}{4 \alpha}(1+ \frac{3}{2\sqrt 2}). \label{displacement}
\end{eqnarray}
Thus, the average speed of the wave packet during the interval $T_{\rm min}$ is
\begin{eqnarray}
\bar v = \frac{\Del r}{ T_{min}}
= \frac{\sqrt 2 +1}{4 \sqrt{2 \pi}}\big|1- \frac{e \theta B}{4\hbar}\big | c \doteq 0.2407 \big |1 - \frac{e \theta B}{4\hbar} \big |c. \label{fr}
\end{eqnarray}

In view of  (\ref{fr}),
we find that our result differs the one in Ref. \cite{vd}. In our result, there is an extra factor $\big |1 - \frac{e \theta B}{4 \hbar} \big |$ which is introduced by the noncommutativity. It is this factor that enables the
the average speed of the wave packet of the electrons to exceed the speed of light in vacuum provided the intensity of the magnetic field is strong enough.

\section{Three-dimensional case}

In this section, we shall generalize our studies in  previous section to three-dimensional case.

For the 3-dimensional case, the noncommutative algebraic relation  is \cite{bertolami2}
\begin{eqnarray}
[\hat x_I, \ \hat x_J ] = i \theta_{IJ}, \quad [\hat p_I, \ \hat p_J ] = i \eta_{IJ}, \no \\
\lbrack \hat x_I, \ \hat p_J] = i \hbar (\del_{IJ} + \frac{\theta_{IK}\eta_{JK}}{4 \hbar^2}),
\label{nps1}
\end{eqnarray}
in which
\begin{eqnarray}
\theta_{IJ} = \eps_{IJK} \theta _K, \quad \eta_{IJ} = \eps_{IJK} \eta_K,
\end{eqnarray}
with $\eps_{IJK}$ being the Levi-Civita symbol. For future purpose, we map the noncommutative variables $\hat x_I, \ \hat p_I$   onto commutative ones by the relation
\begin{equation}
\hat x_I = x_I - \frac{\theta_{IJ}}{2 \hbar}p_J, \ \hat p_I = p_I + \frac{\eta_{IJ}}{2 \hbar} x_J. \label{tdm}
\end{equation}

The Dirac equation in 3-dimensional space takes the same form as  (\ref{2dde}). However, the components $\phi$ and $\chi$ are  two-component spinors, i.e.,
\begin{equation}
\phi= \left( \begin{array}{c} \psi_1 \\ \psi_2 \end{array} \right ), \ \chi =\left( \begin{array}{c} \psi_3 \\ \psi_4 \end{array} \right ),
\end{equation}
and the Hamiltonian (\ref{h10}) becomes
\begin{equation}
H = c \alpha_I( \hat p_I + e \hat A_I) + mc^2 \beta.
\end{equation}
The eigenvalue equation $H \psi = E \psi$ reduces to the form
\begin{eqnarray}
(E- m c^2) \phi = c\sig_I(\hat p_I + e \hat A_I) \chi, \label{eigeneq3} \\
(E+ m c^2) \chi = c\sig_I (\hat p_I + e \hat A_I) \phi. \label{eigeneq4}
\end{eqnarray}
Similar with the 2-dimensional case, $\phi$ is the larger component while the component $\chi$ is the smaller one.
The dynamical equation of the lager component $\phi$ is determined by
\begin{eqnarray}
(E^2 - m ^2 c^4) \phi = c^2 \bigg[( \hat p_I + e \hat A_I)^2
+ i  \eps_{IJK} (\half[ \hat p_I, \ \hat p_J]  \no \\ + \half {e^2}[\hat A_I, \ \hat A_J] + {e}[\hat A_I, \ \hat p_J]) \sig_K \bigg]\phi. \label{tdre}
\end{eqnarray}

We choose the symmetric gauge,
\begin{equation}
\hat A_I = \half \eps_{IJK} B_J \hat x_K, \label{sgtds}
\end{equation}
and align the direction of the magnetic field along the z-axis, i.e., $B_i = B \del_{I3}$. Furthermore, we choose a specific noncommutative configuration
\begin{equation}
\theta_I = \theta \del_{I3}, \ \eta_I = \eta \del_{I3} \label{snc}
\end{equation}
which is widely applied in the studies of noncommutative quantum mechanics.

In this specific noncommutative configuration, the map from the noncommutative variables  to the commutative ones  (\ref{tdm})
becomes rather simple. It is
\begin{eqnarray}
\hat x_i &=& x_i - \frac{\theta}{2 \hbar} \eps_{ij} p_j, \quad \hat p_i = p_i + \frac{\eta}{2 \hbar} \eps_{ij} x_j, \no \\
\hat x_3 &=& x_3, \quad \hat p_3 = p_3.
\end{eqnarray}
Then the equations (\ref{eigeneq3}, \ref{eigeneq4}, \ref{tdre}) turn out to be
\begin{eqnarray}
(E- mc^2) \phi &=& c \big [ (1- \frac{e \theta B}{4 \hbar}) \sig_i p_i \no \\ &&+ (\frac{\eta}{2 \hbar} - \frac{eB}{2}) \eps_{ij} \sig_i x_j + \sig_3 p_3 \big ] \chi, \no \\
(E+ mc ^2) \chi &=& c \big [ (1- \frac{e \theta B}{4 \hbar}) \sig_i p_i \no \\ &&+ (\frac{\eta}{2 \hbar} - \frac{eB}{2}) \eps_{ij} \sig_i x_j + \sig_3 p_3 \big] \phi \label{eigeneq6}
\end{eqnarray}
and
\begin{eqnarray}
(E^2 - m^2 c^4)  \phi &=& c^2 \bigg[ (1- \frac{e \theta B}{4 \hbar})^2  p_i^2 + p_3^2+ (\frac{\eta}{2 \hbar} - \frac{eB}{2})^2 x_i ^2 \no \\ && - 2 (\frac{\eta}{2 \hbar} - \frac{e B}{2})(1- \frac{e\theta B}{4 \hbar})L_z)  \no \\
&&+ 2 \hbar (\frac{eB}{2}- \frac{\eta}{2 \hbar}  )(1 - \frac{e \theta B}{4 \hbar}) \sig_3 \bigg] \phi.
\end{eqnarray}

Compared with the 2-dimensional case (\ref{cde}), we find that besides the free motion in the $z$ direction, there is an explicit spin-magnetic  coupling term. Since there are no orbit-spin coupling, the structure of the 3-dimensional case is very similar with the 2-dimensional one. For simplicity, we assume that the initial states  have $+ \half$ spin projection along the $z$ axis (we label it as $| \uparrow \rangle$) and a Gaussian packet in the same direction \cite{vd}
\begin{equation}
\psi_3 (z) = \frac{1}{{(2 \pi d^2)}^{1/4}}  e ^{{-z^2}/{4d^2}} e ^{ip z}.
\end{equation}
The eigenvalues of the above equation can be achieved easily. The eigenvalues are
\begin{equation}
E_{n,m_l} = \pm \sqrt {m^2 c^4 + c^2 p ^2 + 2 m c^2 \hbar \Omega (n+m_l +2)}, \\
\end{equation}
where $\Omega$ has been given in (\ref{frequency}).

In order to make a comparison with Ref. \cite{vd}, we choose the larger components of two initial states
\begin{eqnarray}
\phi = \left( \begin{array}{c} \psi_1 \\ \psi_2 \end{array} \right) = \phi_{0,0} \otimes | \uparrow \rangle
\ {\rm and} \ \phi =  \left( \begin{array}{c} \psi_1 \\ \psi_2 \end{array} \right)= \phi_{2,0} \otimes | \uparrow \rangle.
\end{eqnarray}
The corresponding smaller components can be calculated by using (\ref{eigeneq6}). They are
\begin{equation}
\chi=  \left( \begin{array}{c} \psi_3 \\ \psi_4 \end{array} \right) = \chi_{0,0} \otimes| \uparrow \rangle \ {\rm and} \ \chi =  \left( \begin{array}{c} \psi_3 \\ \psi_4 \end{array} \right)= \chi_{2,0} \otimes| \uparrow \rangle.
\end{equation}
Thus, two steady states we prepared are
\begin{equation}
\Phi_{0,0} = N_{0,0}\left (\begin{array}{c}\phi_{0,0} \\ 0 \\ \frac{cp \phi_{0,0}}{E_{0,0} + m c^2} \\ \frac{2i \hbar \alpha c (1 - \frac{e \theta B}{4 \hbar}) \phi_{1,1}}{E_{0,0} + m c^2} \end{array} \right ) \psi_3(z) e^{-i \frac{E_{0,0}t}{\hbar}}
\end{equation}
and
\begin{equation}
\Phi_{2,0} = N_{2,0}\left (\begin{array}{c}\phi_{2,0} \\ 0 \\ \frac{cp \phi_{2,0}}{E_{2,0} + m c^2} \\ \frac{2 \sqrt 2 i \hbar \alpha c (1 - \frac{e \theta B}{4 \hbar}) \phi_{3,1}}{E_{2,0} + m c^2} \end{array} \right ) \psi_3 (z) e^{-i \frac{E_{2,0}t}{\hbar}},
\end{equation}
where
\begin{equation}
N_{0,0} = \frac{E_{0,0} + mc^2}{\sqrt{(E_{0,0}+ m c^2)^2 + c^2 p^2 + 4 \hbar^2 \alpha^2 c^2 (1 - \frac{e \theta B}{4 \hbar})^2}}
\end{equation}
and
\begin{equation}
N_{2,0} = \frac{E_{2,0} + mc^2}{\sqrt{(E_{2,0}+ m c^2)^2 + c^2 p^2 + 8 \hbar^2 \alpha^2 c^2 (1 - \frac{e \theta B}{4 \hbar})^2}}
\end{equation}
are two normalization constants.

Taking the limit of $B \to \infty$, we find that the eigenvalues of the two eigenfunctions and the normalization constants are
\begin{equation}
E_{0,0} = 2 \big |1 - \frac{e \theta B}{4\hbar} \big | \alpha c, \quad E_{2,0}  = 2 \sqrt 2 \big |1 - \frac{e \theta B}{4\hbar} \big| \alpha c
\end{equation}
and
\begin{equation}
N_{0,0}= N_{2,0} = \frac{1}{\sqrt 2}.
\end{equation}
We superpose the two steady states homogeneously as our initial state
\begin{equation}
\Psi (r, \varphi, z, t)=\frac{1}{\sqrt 2} [\Phi_{0,0}(r, \varphi, z, t) + \Phi_{2,0}(r, \varphi, z, t)].
\end{equation}
Then the minimum time for this  state evolving from the initial state $\Psi(r, \varphi, z, 0)$ to the orthogonal state $\Psi(r, \varphi, z, T_{\rm min})$ can be calculated directly. It is equivalent to the one we get in (\ref{mt}). And the displacement along the radial direction of the wave packet during this period time is nothing but the one in (\ref{displacement}). Since the minimum time and the displacement of the wave packet are all same as the two-dimensional case, we can draw the conclusion that it is possible for the average speed of the electron's wave packet to exceed the speed of light in vacuum. 

\section{Conclusions and Remarks}

In this paper, we study the influence of the noncommutativity on the average speed of the wave packet of an electron along the radial direction,  both 2-dimensional and 3-dimensional cases are investigated. In fact, our paper is the noncommutative generalization of Ref. \cite{vd}.  For 3-dimensional case, we choose a specific noncommutative configuration (\ref{snc})
which reduces the 3-dimensional noncommutative quantum model to a 2-dimensional noncommutative one and a free motion along the $z$-direction. With the help of the exact solutions of the 2-dimensional model, one can resolve the 3-dimensional case easily. Although the choice of the noncommutative configuration (\ref{snc}) may lose some generalities, it is enough for our purpose.

We find that due to the noncommutativity, it is possible for the wave packet of an electron to travel
in a speed faster than  light in  vacuum provided the intensity of the magnetic field is strong enough. It  obviously  conflicts with the special relativity. Therefore, we find a
clear evidence of violating  Lorentz invariance in the relativistic quantum mechanics region.

It should be emphasized that there is a fundamental difference between our result and the one in Refs. \cite{jackiw, cai}. In these references, the authors show that because of the noncommutativity, the speed of the electromagnetic wave, which is massless from the quantum point of view, is different from the assumption of special relativity.  Our results show that due to the noncommutativity, even massive particles, namely, electrons, can travel faster than light in the vacuum.
Therefore, besides the noncommutative QED, we find a clear signature of violating the special relativity in the quantum mechanics region due to the noncommutativity.

\section*{Acknowledgement}
This work is supported by the NSFC with Grant No 11465006.

\end{document}